\documentclass[9pt,twocolumn,twoside]{opticajnl}
\journal{opticajournal} 

\setboolean{shortarticle}{true}

\usepackage{lineno}

\title{Ghost imaging-based Non-contact Heart Rate Detection}

\author[1]{Jianming Yu}
\author[1,*]{Yuchen He}
\author[2,*]{Bin Li}
\author[1]{Hui Chen}
\author[1]{Huaibin Zheng}
\author[1]{Jianbin Liu}
\author[1]{Zhuo Xu}

\affil[1]{Electronic Materials Research Laboratory, Key Laboratory of the Ministry of Education and International Center for Dielectric Research, School of Electronic Science and Engineering, Xi'an Jiaotong University, Xi'an 710049, China}
\affil[2]{Bioinspired Engineering and Biomechanics Center (BEBC), Xi’an Jiaotong University, Xi’an 710049, China}

\affil[*]{yuchenhe@xjtu.edu.cn; binli@xjtu.edu.cn}

\begin{abstract}
 Remote heart rate measurement is an increasingly concerned research field, usually using remote photoplethysmography (rPPG) to collect heart rate information through video data collection. However, in certain specific scenarios (such as low light conditions, intense lighting, and non-line-of-sight situations), traditional imaging methods fail to capture image information effectively, that may lead to difficulty or inability in measuring heart rate. To address these limitations, this study proposes using ghost imaging as a substitute for traditional imaging in the aforementioned scenarios. The mean absolute error between experimental measurements and reference true values is 4.24 bpm.Additionally, the bucket signals obtained by the ghost imaging system can be directly processed using digital signal processing techniques, thereby enhancing personal privacy protection.
\end{abstract}

\setboolean{displaycopyright}{false} 

\begin{document}

\maketitle

\section{Introduction}
Remote photoplethysmography (rPPG) is a technique that utilizes sensors, such as cameras, to capture the periodic changes in skin color induced by the cardiac cycle\cite{Verkruysse:08,7319857}. rPPG technology is commonly utilized for extracting blood volume pulses (BVP) and measuring physiological parameters related to cardiac circulation, encompassing heart rate, respiratory rate, and heartbeat variability\cite{10.1063/5.0043865,10.1117/1.JBO.18.6.061205}. Physiological indicators like heart rate data can also serve specific purposes such as crime detection. In recent years, significant advancements have been achieved in the evaluation and measurement of physiological indicators based on rPPG, leading to a proliferation of diverse algorithms aimed at enhancing the robustness of rPPG\cite{10208697,7873222,8756117,9484675}.

However, the majority of existing research in this field heavily rely on video image inputs acquired using conventional cameras, which introduces a high susceptibility to environmental factors during the signal extraction process. Specifically, in challenging scenarios where traditional imaging methods fail to effectively capture information, such as in low-light conditions\cite{9320298}, intense illumination, and non-line-of-sight situations, the applicability of rPPG technology encounters significant limitations and challenges\cite{10.1117/1.JBO.21.11.117001,7565547,9913818}.
 
The imaging effect of ghost imaging(GI) surpasses that of traditional imaging in the aforementioned scenarios. GI employs an active lighting method, which differs from the passive lighting method utilized in traditional imaging\cite{shapiro2008computational,ferri2005high}. Additionally, numerous studies have been conducted on imaging within these scenarios in the field of GI. Therefore, this paper proposes the utilization of GI as a substitute for conventional imaging and rPPG technology in order to extract heart rate, thereby addressing the limitations of existing rPPG technology in special environments\cite{Jiang:23, 10.1063/5.0194784, Yang:20, Li_2020}. Furthermore, this study also investigates the direct extraction of heart rate from the bucket signal utilized in GI. The direct extraction of heart rate from the bucket signal obviates the need for imaging, thereby enhancing privacy protection.

\section{Method}
The periodic pulsation of the human heart induces dynamic changes in the volume of blood vessels beneath the skin, resulting in changes in skin surface color. Following the dichromatic model\cite{7565547}, the reflection of skin pixels captured within the recorded image sequence can be defined as a time-varying function:
\begin{equation}
C(t) = I(t)\cdot(v_{s}(t)+v_{d}(t))+v_{n}(t).
\label{eq1}
\end{equation}

Where $I(t)$ represents the temporal variation in illumination intensity, $v_s (t)$ denotes specular reflection, $v_d (t)$ refers to diffuse reflection, and $v_n (t)$ accounts for noise in the measurement process. The rPPG signal is embedded within the diffuse reflection.Therefore, it may be feasible to replace traditional imaging systems with ghost imaging systems.

\begin{figure*}[ht]
\centering
\fbox{\includegraphics[width=15cm, height = 4.5cm]{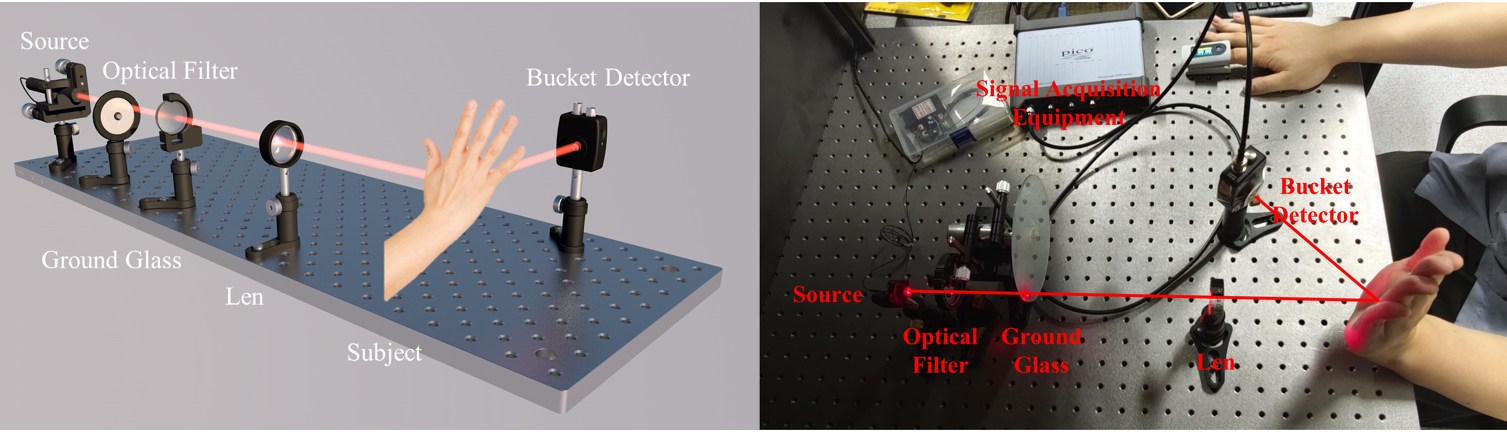}}
\caption{Replacement of traditional imaging methods by GI.The left is the schematic diagram, and the right is the experimental set.}
\label{Fig. 1}
\end{figure*}

The proposed method, illustrated in Fig.\ref{Fig. 1}, involves utilizing a ghost imaging system to acquire information about the target area of the subject. Heart rate can be obtained either by restoring the image or, alternatively, without undergoing the image reconstruction process, directly through digital signal processing on the bucket signals.

The algorithm for reconstructing GI images, using differential ghost imaging (DGI)\cite{PhysRevLett.104.253603} as an illustrative example, can be formulated as follows:
\begin{equation}
G = <I\cdot B> (1 - \frac{<B>} {<B'>}) .
\label{eq2}
\end{equation}

Where $I$ represents speckle, $B$ denotes the bucket detection signal of the object irradiated by speckle, and $B'$ refers to the bucket detection signals of speckles. In the GI system, the bucket detection signals of the measured object corresponds to a time-varying function $C(t)$ as described in equation (\ref{eq1}), albeit with speckle modulation affecting the light intensity term. The bucket signal can be expressed as the following equation.
\begin{equation}
B(t) = C(t) = I(t)\cdot(v_{s}(t)+v_{d}(t)+v_{n}(t)).
\label{eq3}
\end{equation}

As the bucket detector directly detects light intensity, the noise from the bucket detector and signal acquisition equipment is also modulated by the light intensity term.Consequently, the diffuse reflection information $v_d (t)$, which encompasses the rPPG signals, is solely modulated in terms of intensity while preserving the integrity of vibration information, it remains feasible to directly extract the heart rate from the bucket detection signals of the measured object. After obtaining the rPPG signals, heart rate estimation can be achieved through time-frequency domain analysis. The most commonly utilized approach for heart rate calculation involves performing a Fourier transform on the filtered rPPG signals, finding its peak value in the frequency domain, and subsequently computing it:

\begin{equation}
Hr=60\cdot f_{max}.
\label{eq4}
\end{equation}

Where $Hr$ represents the heart rate value, and $f_{max}$ represents the maximum peak value in the frequency domain.

 The Mean Absolute Error (MAE) has been used for performance evaluation of GI-based heart rate detection methods. It represents the average of the absolute error between the true value and the observed value. The calculation formula is as follows:

 \begin{equation}
MAE=\frac{1}{N}\sum_{i=1}^{n}|Y_i-X_i| .
\label{eq5}
\end{equation}

Where $Y_i$ and $X_i$ represent the reference measured value and true value, respectively, for the ith value, and N is the total number of values.

In processing the bucket detection signals obtained from actual experiments, we employed Variational Mode Decomposition (VMD)\cite{6655981}, and its constrained variational model can be represented as:

\begin{equation}
     \begin{aligned}
    \min_{\{u_{k}\},\{\omega_{k}\}}\left\{\sum_{k}\left\Vert\partial_{t}\left[\left(\delta(t)+{\frac{j}{\pi t}}\right)\ast u_{k}(t)\right]e^{-j\omega_{k}t}
    \right\Vert_{2}^{2}\right\} \\
    \hfill{\rm s.t.}\quad\sum_{k}u_{k}=f.
    \end{aligned}
\label{eq6}
\end{equation}

Where $\{u_k\}$ and $\{\omega k\}$ are shorthand notations for the set of all modes and their center frequencies, respectively.

\section{Simulation and Experiment}

\begin{figure}[ht]
\centering
\fbox{\includegraphics[width=\linewidth]{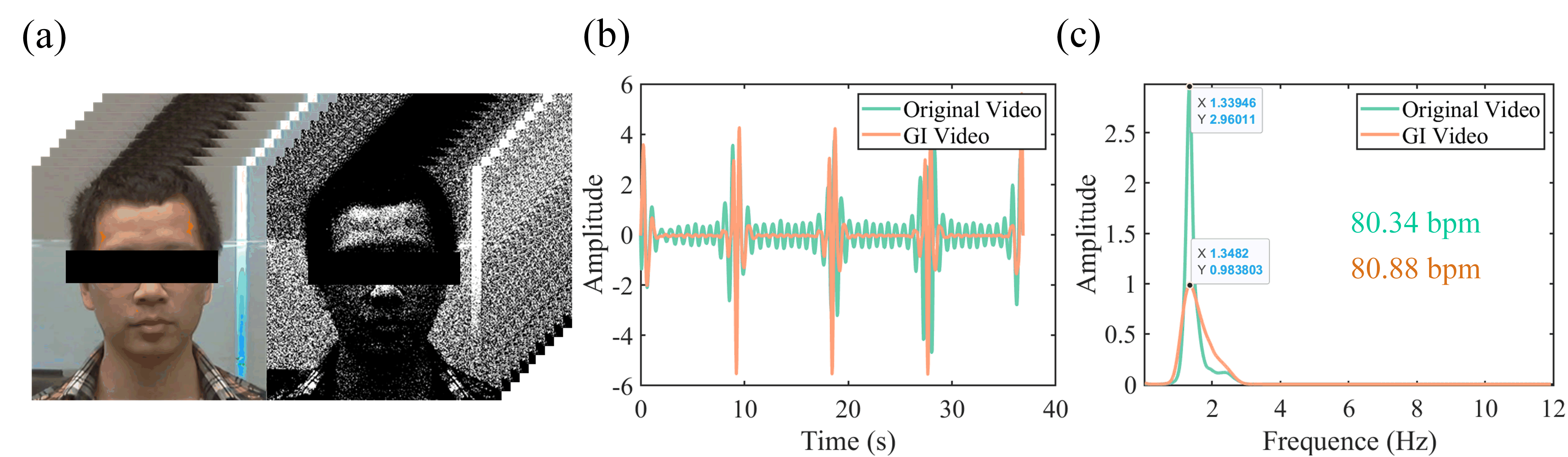}}
\caption{(a) Original video frames (224*224, 24fps) and video frames of DGI simulation. (b)  rPPG signals after filtering. (c) Fourier transform result of rPPG signals.}
\label{Fig. 2}
\end{figure}

Firstly, we conducted simulation validation based on GI-reconstructed images. The images were reconstructed using DGI, followed by inputting the image information into a trained neural network to achieve results comparable to those obtained through conventional imaging techniques. The images utilized in the simulation are sourced from the VIPL-HR\cite{niu2018viplhr} dataset, while the used neural network model is PhysFormer\cite{Yu_2022_CVPR}. The result is illustrated in Fig. \ref{Fig. 2}.

The rPPG signals, in Fig. \ref{Fig. 2} (b), were generated by the PhysFormer model based on the original video and DGI video respectively, and they had been filtered through a hamming window bandpass filter from 0.7 Hz to 4 Hz. What is shown in Fig. \ref{Fig. 2}(c) is the spectrum obtained by the fourier transform of the two rPPG signals in Fig. \ref{Fig. 2}(b), where the peak value of the rPPG signal obtained from the DGI video is 1.348Hz, and the peak value of the rPPG signal obtained from the original video is 1.339Hz. After calculation, the heart rate values can be obtained, 80.9 bpm and 80.3 bpm respectively, and the ground truth is 83.6 bpm.

\begin{figure}[ht]
\centering
\fbox{\includegraphics[width=\linewidth]{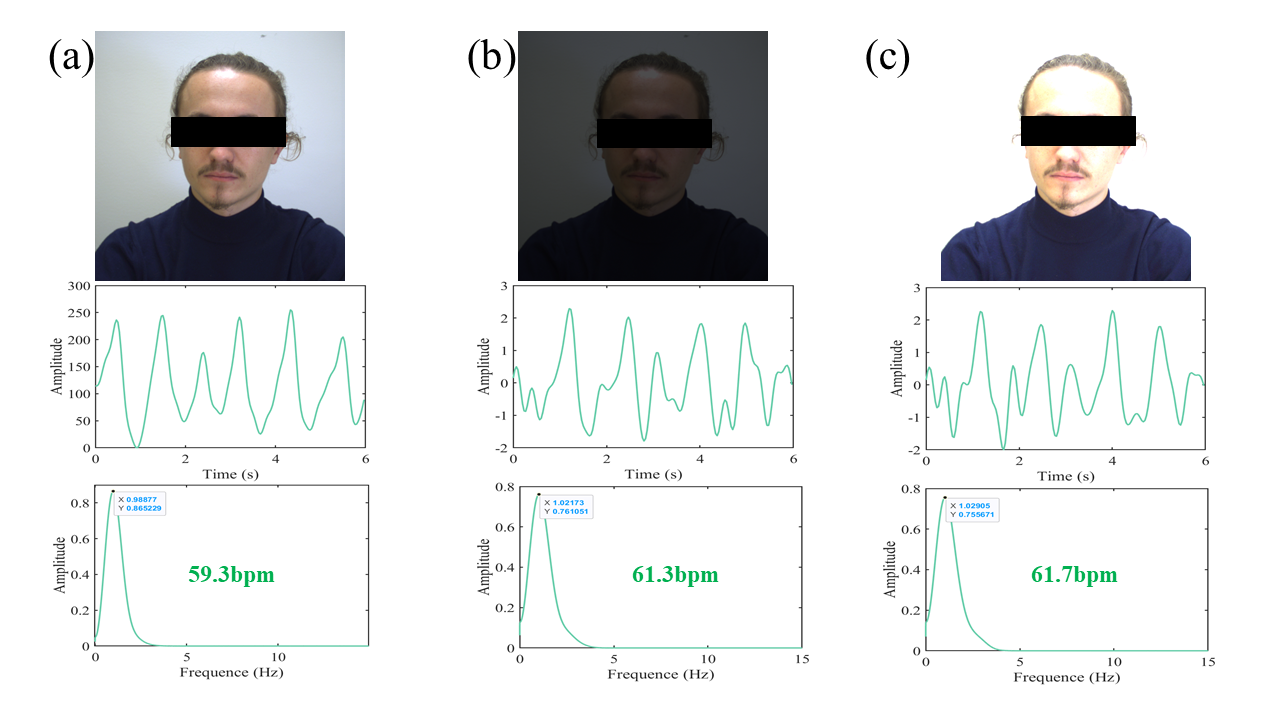}}
\caption{Measurement results simulated in different lighting environments (a) Normal illumination. (b) Low light environment. (c) Intense light environment.}
\label{Fig. 3}
\end{figure}

In this work, simulation tests were conducted to extract heart rate solely from raw bucket signals using the UBFC-Phys\cite{9346017} dataset. A subset of videos from the UBFC-Phys dataset was chosen and divided into 6-second segments to simulate the extraction of instantaneous heart rate through rPPG signals. During the simulation, the green channel of the images was utilized as a proxy for the rPPG signal, simulating random speckle illumination of the skin, mimicking the use of green light (500-560nm) to illuminate the skin. According to the principles of rPPG technology, any area on the skin surface can be selected for illumination\cite{TAKANO2007853,Allen_2007}. For this particular simulation, we chose the forehead region, and an example result is presented in Fig. \ref{Fig. 3}(a). The rPPG signal extracted from the bucket signal exhibited a frequency domain peak of 0.989Hz, while the original BVP data provided in the UBFC-Phys dataset showed a frequency domain peak of 0.977Hz. The calculated instantaneous heart rates were 59.3bpm and 58.6bpm, respectively. The heart rate values measured across the 30 video segments in the simulation demonstrated high consistency with those derived from the BVP data provided by the UBFC-Phys dataset.

Furthermore, this research also carried out simple simulations under dim and bright light conditions, as depicted in Fig. \ref{Fig. 3}(b) and Fig. \ref{Fig. 3}(c), respectively. Under these simulated special environments, the rPPG waveform exhibited noticeable changes, yet the frequency domain peaks remained relatively stable. Specifically, the frequency domain peaks were 1.022 Hz and 1.029 Hz in the dim and bright light conditions, respectively, corresponding to calculated heart rates of 61.3 bpm and 61.7 bpm.

\begin{figure}[ht]
\centering
\fbox{\includegraphics[width= 8cm]{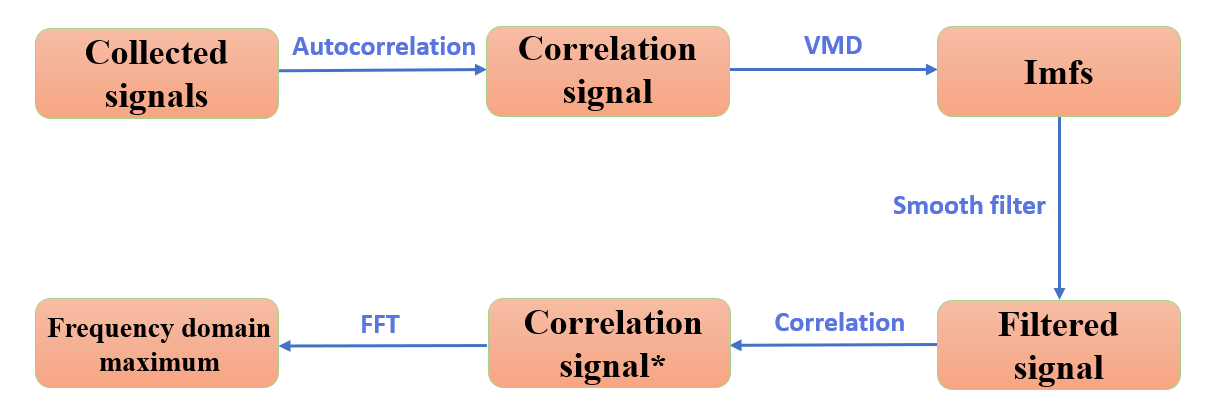}}
\caption{Algorithm flow chart.}
\label{Fig. 4}
\end{figure}

In addition, a key aspect of this work involves directly extracting heart rate information from bucket signals within the GI system without reconstructing an image, a part which has been both simulated and experimentally validated, albeit under more complex circumstances in the experimental setting.
In practical experiments, there exists a background noise issue, and the heart rate signal is a very weak periodic signal (which can be approximated as a sine wave), often submerged beneath the noise level. Therefore, in response to the problem of system noise floor, a method has been proposed for extracting the period of weak signals under low signal-to-noise ratio (SNR) conditions, as illustrated in Fig. \ref{Fig. 4}. The experimental setup for extracting heart rate directly from the bucket signal is shown on the right side of Fig. \ref{Fig. 1}.

\begin{figure}[ht]
\centering
\fbox{\includegraphics[width= 7cm]{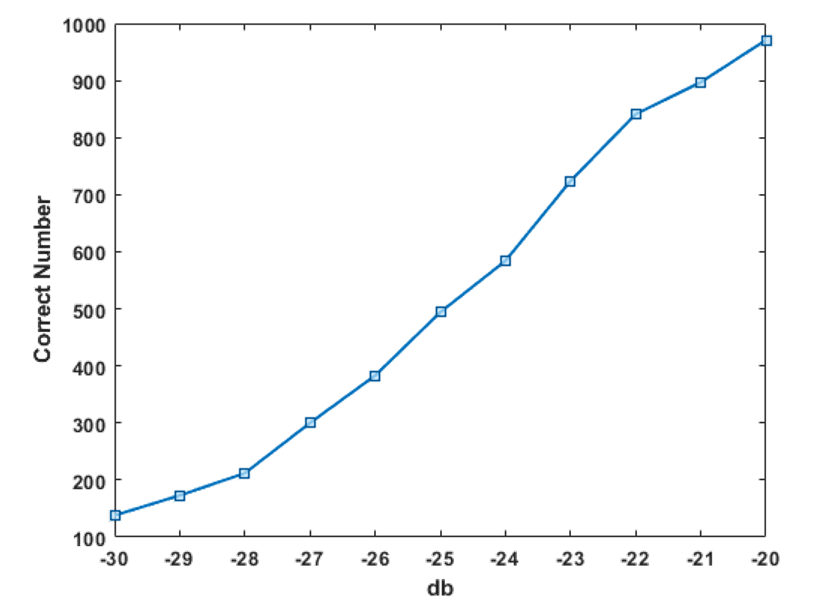}}
\caption{The performance of the algorithm under different noises.}
\label{Fig. 5}
\end{figure}

This method requires prior knowledge of the frequency range of the periodic signal, which for heart rate signals typically falls within the range of 0.7 Hz to 4 Hz.
After testing, it was discovered that under conditions allowing for ±0.02 deviations in peak frequency domain values, this method can effectively extract signal periods in an additive white Gaussian noise environment with an SNR of -20 dB, achieving a success rate of 97.1\%. However, when the additive white Gaussian noise decreases to -25 dB, the success rate drops to 49.4\%, with specific details shown in Fig. \ref{Fig. 5}.
Due to the presence of system noise that prevents one hundred percent successful determination of the heart rate signal's period, data that are closer to the actual values are selected as valid data. Despite the effectiveness of the method at certain noise levels, its performance diminishes at lower noise intensities, highlighting the ongoing challenge of accurate signal extraction in noisy environments.

\begin{figure*}[ht]
\centering
\fbox{\includegraphics[width=14cm, height = 6cm]{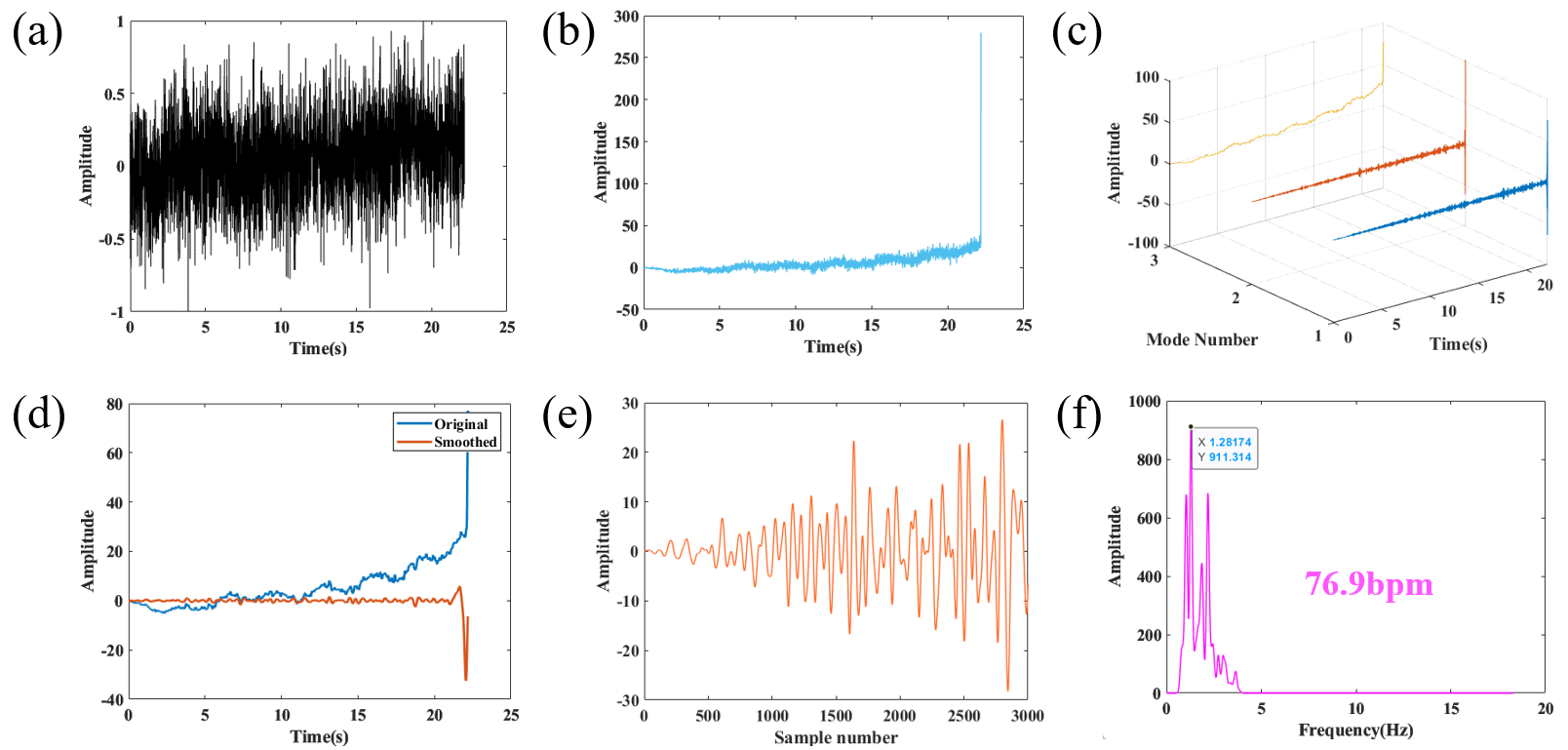}}
\caption{Data processing example (a)Original bucket signal.(b)Bucket signal autocorrelation(first half part).(c)VMD result.(d)The selected imf before and after filtering.(e)The cross-correlation function between filtered imf and the original signal(first half part).(f)Frequency domain of the cross-correlation function.}
\label{Fig. 6}
\end{figure*}

The data with a calculated heart rate cycle that differs from the actual value by no more than 10 were considered valid data, resulting in a total of 50 sets of data were saved. The mean absolute error(MAE) for these data was found to be 4.24bpm. The proposed method for extracting heart rate from the acquired bucket signals is illustrated in Fig. \ref{Fig. 6}. 

The autocorrelation function's length being twice that of the original signal, so Fig. \ref{Fig. 6}(b) presents the first half of the autocorrelation of the barrel signals. Fig. \ref{Fig. 6}(c) displays the three intrinsic mode functions (imfs) obtained through VMD, from which the one with the lowest frequency is selected and filtered to yield Fig. \ref{Fig. 6}(d). Finally, by performing cross-correlation on the filtered signal with the raw barrel signals and determining the peak in the frequency domain, the heart rate value can be calculated as 76.9bpm using equation (\ref{eq4}), the difference from the actual value is 1.1bpm..

\section{Conclusion}
In summary, using the GI architecture has effectively realized remote heart rate measurement and proposed a method for extracting the periodicity of weak signals from systems with low SNR, demonstrating promising performance under challenging environmental conditions. Moreover, obtaining heart rate directly from the bucket signal enhances privacy protection. However, due to the system's SNR limitations, it cannot guarantee that every measurement yields usable data. Substantial improvements to the practicality of this system can be achieved through refinements in the experimental setup and further research into methods for extracting faint signals.

\section{Funding}
This work was supported in part by the National Natural Science Foundation of China under Grant 62375215, National Funding Postdoctoral Researcher Program of China under Grant GZC20232102, 2024QCY-KXJ-183HZ, and in part by the 111 Project of China underGrant B14040. 

\section{Disclosures}
The authors declare no conflicts of interest.

\section{Data availability}
Data underlying the results presented in this paper are not publicly available at this time but may be obtained from the authors upon reasonable request.




\end{document}